\documentstyle[aps,prb,preprint]{revtex}
\input psfig.tex
\begin{document}
\author{J. P. Neirotti \\
Department of Chemistry, University of Rhode Island\\
51 Lower College Road, Kingston, RI 02881-0809\\$ $
\\
F. Calvo \\ D\'{e}partement de Recherche sur la Mati\`{e}re
Condens\'{e}e \\
Service des Ions, Atomes et Agr\'egats\\
CEA Grenoble \\ F38054 Grenoble Cedex, France
\\ $ $ \\
David L. Freeman \\ Department of Chemistry, University of Rhode
Island\\
51 Lower College Road, Kingston, RI 02881-0809\\
and\\
J. D. Doll\\
Department of Chemistry, Brown University\\
Providence, RI 02912}
\title{Phase changes in 38 atom
Lennard-Jones clusters. I: A parallel tempering study in the canonical
ensemble}
\date{\today}
\maketitle

\begin{abstract}
The heat capacity and isomer distributions of the 38 atom Lennard-Jones
cluster have been calculated in the canonical ensemble using parallel
tempering Monte Carlo methods.  A distinct region of temperature is
identified that corresponds to equilibrium between the global minimum
structure and the icosahedral basin of structures.  This region of
temperatures occurs below the melting peak of the heat capacity and is
accompanied by a peak in the derivative of the heat capacity with
temperature.  Parallel tempering is shown to introduce correlations
between results at different temperatures.  A discussion is given that
compares parallel tempering with other related approaches that ensure
ergodic simulations.
\end{abstract}
\pacs{{\bf PACS} numbers: }

\section{Introduction}

Because the properties of molecular aggregates
impact diverse areas ranging from nucleation and
condensation\cite{abraham} to heterogeneous catalysis, the study of
clusters 
has 
continued to be an important part of modern condensed matter science. 
Clusters can be viewed as an intermediate phase of matter,
and clusters can provide
information about the transformation from finite to bulk behavior. 
Furthermore, the potential surfaces of clusters can be complex, and
many clusters are useful prototypes for studying
other systems having complex phenomenology.

The properties of small
clusters can be unusual owing to the dominance of surface rather than
bulk atoms.  A particularly important and well studied example of a
property that owes its behavior to the presence of large numbers of
surface atoms is cluster structure.\cite{wd,fd,xie} 
 The structure of clusters can differ
significantly from the structure of the corresponding bulk material, and
these differences in structure have implications about the properties of
the clusters.  For example, most small Lennard-Jones (LJ) clusters have
global potential surface minima that are based on icosahedral growth
patterns.  The five-fold symmetries of these clusters differ
substantially from the closest-packed arrangements observed in bulk
materials.

While most small Lennard-Jones clusters have geometries based on
icosahedral core structures, there can be
exceptions.\cite{wd,dmw,dmw1,ld}  
A notable
example is 
the 38-atom Lennard-Jones cluster [LJ$_{38}$]. This cluster is
particularly 
interesting
owing to its complex potential surface and associated phenomenology. 
The potential surface for LJ$_{38}$ has been described in detail by
Doye, Miller and Wales\cite{dmw} 
who have carefully constructed the disconnectivity
graph\cite{ce,bk} for the system using information garnered from basin
hopping
and eigenvector following
studies of the low energy potential minima along with examinations of
the transition state barriers.  
The general structure of this potential surface can be
imagined to be two basins of similar energies separated by a large
energy
barrier with the lowest energy basin being significantly narrower than
the second basin.
Striking is the global minimum energy
structure for LJ$_{38}$ which, unlike the case for most small
Lennard-Jones clusters, is not based on an icosahedral core, but rather
is a symmetric truncated octahedron.  The vertices defined by the
surface atoms of LJ$_{38}$ have a morphology
identical to the first Brillouin zone of a face centered cubic 
lattice,\cite{kittel} and the high
symmetry of the cluster may account for its stability.  It is
interesting to note that recent experimental studies\cite{parks} 
of nickel clusters 
using nitrogen
uptake measurements have found the global minimum
of Ni$_{38}$ to be a truncated octahedron as well.
The basin of energy minima about
the global minimum of LJ$_{38}$ is narrow compared to the basin about
the next highest energy isomer which does have an icosahedral core.  The
difference in energy between the global minimum and the lowest minimum
in the icosahedral basin is only 0.38\%
of the energy of the global minimum.\cite{dmw}

Characteristic of some thermodynamic properties of small clusters are
ranges of temperature over which these properties change rapidly in a
fashion reminiscent of the divergent behavior known to occur in bulk
phase transitions at a single temperature.  The rapid 
changes in such thermodynamic properties
for clusters are not divergent and occur over a range of temperatures
owing to the finite sizes of the systems.
In accord with the usage introduced by Berry, Beck, Davis and
Jellinek\cite{berry}
we refer to the temperature ranges where rapid changes occur as ``phase
change'' regions, rather than using the term ``phase transition,'' that
is reserved
for systems at the thermodynamic limit.
As an example LJ$_{55}$ displays a heat capacity
anomaly over a range in temperatures often associated with what has been
termed ``cluster melting.''\cite{labwhet} 
Molecular dynamics and microcanonical
simulations performed at kinetic temperatures in
the melting region of LJ$_{55}$
exhibit van der Waals type
loops in the caloric curves and coexistence between solid-like and
liquid-like forms.

In recent studies, Doye, Wales and Miller\cite{dwm} 
and Miller, Doye and Wales\cite{mdw}
have examined the
phase change behavior of LJ$_{38}$.  These authors have 
calculated the heat
capacity and isomer distributions as a function of temperature using
the superposition method.\cite{sup1,sup2}  In the superposition method
the microcanonical density of states is calculated for each potential
minimum, and the total density of states is then
constructed by summation with respect to each local density of states. 
Because it is not possible to find all potential minima for a system as
complex as LJ$_{38}$, the summation is augmented with factors that
represent the effective weights of the potential minima that are
included in the sum.
The superposition method has also been improved to account for
anharmonicities and stationary points.\cite{sup2}
For LJ$_{38}$ Doye {\em et al.}\cite{dwm} have identified
two phase change regions.  The first, accompanied by a heat
capacity maximum, is associated with a solid-to-solid phase change
between the truncated octahedral basin and the icosahedral basin.  A
higher temperature heat capacity anomaly represents the solid-liquid
coexistence region, similar to that found in other cluster systems. The
heat
capacity anomaly associated with the melting transition in LJ$_{38}$ is
steeper and more pronounced than the heat capacity peak 
that Doye {\em et al.}\cite{dwm} have associated with the
solid-solid transition.
Because the weights that enter in the sum to construct the
microcanonical density of states are estimated, it is important to
confirm the findings of Doye {\em et al.}\cite{dwm} by detailed
numerical
simulation.  Such simulations are a goal of the current work
and its companion paper.  As is
found in Section III, the simulations provide a heat
capacity curve for LJ$_{38}$ that has some qualitative differences with 
the curve reported by Doye {\em et al.}\cite{dwm}

Owing to the complex structure of the potential surface of LJ$_{38}$,
the system represents a particularly challenging case for simulation. 
It is well known that simulations of systems having more than one
important region of space separated by significant energy barriers can
be difficult.  The difficulties are particularly severe if any of the
regions are either narrow or reachable only via narrow channels.  The
narrow basin about the global minimum makes simulations of LJ$_{38}$
especially difficult.
There are several methods that have been developed that can prove to be
useful in overcoming such ergodicity difficulties in simulations.  Many
of these methods use information about the underlying potential surface
generated from simulations on the system using parameters where the
various regions of configuration space are well-connected.  One of the
earliest of these methods is J-walking\cite{jwalk1} 
where information about the
potential surface is obtained from simulations at high temperatures, and
the information is passed to low temperature walks by jumping
periodically to the high temperature walk.  Closely allied with
J-walking is the parallel tempering
method\cite{temper1,temper2,temper3,temper4,temper5}
where configurations are
exchanged between walkers running at two differing temperatures. 
A related approach,\cite{straub} similar in spirit to J-walking, uses
Tsallis distributions that are sufficiently broad to cover much of
configuration space.  Another recent addition\cite{berne} to these
methods
is the use of multicanonical distributions\cite{guber} in the jumping
process. 
Multicanonical walks are performed using the entropy of the
system, and multicanonical distributions are nearly independent of the
energy thereby allowing easy transitions between energy basins.  As we
discuss in the current work, we have found the parallel tempering
method to be most useful in the context of simulations of LJ$_{38}$. 
A comparative discussion of some of 
the methods outlined above is given later in
this paper.

In the current work we apply parallel tempering to the calculation of
the thermodynamic properties of LJ$_{38}$ in the canonical ensemble.  In
the paper that follows\cite{paper2} we again use parallel tempering to
study
LJ$_{38}$, but using molecular dynamics methods along with
microcanonical Monte Carlo simulations.
Our goals are to understand better
this complex system and to determine the best simulation
method for systems of comparable complexity.  The contents of the
remainder of this first paper are as follows.  In Section II we discuss
the
methods used with particular emphasis on the parallel tempering
approach and its relation to the J-walking method.
In Section III we present the results including the heat
capacity as a function of temperature and identify the phase change
behaviors of LJ$_{38}$.  In Section IV we present our conclusions and
describe our experiences with alternatives to parallel tempering for
insuring ergodicity.

\section{Method}

For canonical simulations we model a cluster with $N$ atoms 
by the standard Lennard-Jones
potential augmented by a constraining potential $U_c$ used to define the
cluster

\begin{equation}
U({\bf r})=4 \varepsilon \sum_{i<j}^N \left [ \left (
\frac{\sigma}{r_{ij}} \right )^{12}
-\left ( \frac{\sigma}{r_{ij}} \right )^6 \right ] +U_c,
\end{equation}
where $\sigma$ and $\varepsilon$ are respectively the standard
Lennard-Jones length
and energy parameters, and $r_{ij}$ is the distance between particles
$i$ and 
$j$.
The constraining potential is necessary because clusters at defined
temperatures
have finite vapor pressures, and the evaporation events can make the
association
of any atom with the cluster ambiguous.  For classical Monte Carlo
simulations,
a perfectly reflecting constraining potential is most convenient
\begin{equation}
U_c=\sum_{i=1}^N u({\bf r}_i),
\end{equation}
with
\begin{equation}
u({\bf r})=\left\{ \begin{array}{ll}
\infty&|{\bf r}-{\bf r}_{cm}|>R_c \\
0& |{\bf r}-{\bf r}_{cm}|<R_c \\
\end{array}
\right.
\end{equation}
where ${\bf r}_{cm}$ is the center
of mass of the cluster, and we call $R_c$ the constraining radius.

Thermodynamic properties of the system are calculated with Monte Carlo
methods
using the parallel tempering
technique.\cite{temper1,temper2,temper3,temper4,temper5}
To understand the application of the
parallel tempering method and to understand the comparison of parallel
tempering
with other related methods,
it is useful to review the basic principles
of Monte Carlo simulations.  

In the canonical ensemble the goal is the calculation of
canonical expectation values.  For example, the average potential energy
is
expressed
\begin{equation}
\langle U \rangle = \frac{\int d^{3N} r \ U({\bf r}) e^{-\beta
U({\bf r})}}{\int d^{3N} r \ e^{-\beta U({\bf r})}},
\end{equation}
where $\beta=1/k_BT$ with $T$ the temperature and $k_B$ the Boltzmann
constant.
In Monte Carlo simulations such canonical averages are determined by
executing a
random walk in configuration space so that the walker visits points in
space
with a probability
proportional to the canonical density $\rho({\bf r})=
Z^{-1}\exp[-\beta
U({\bf r})]$, where $Z$ is the configurational integral
that normalizes the density.
After generating $M$ such configurations in a random
walk, the expectation value of the potential energy is approximated by
\begin{equation}
\langle U \rangle_M =\frac{1}{M}\sum_{i=1}^M U({\bf r}_i).
\end{equation}
The approximate expectation value $\langle U \rangle_M$ becomes exact in
the
limit that $M \rightarrow \infty$.  

A sufficiency condition for the random walk
to visit configuration space 
with a probability proportional to the density $\rho({\bf r})$ is the 
detailed balance condition\cite{kalos,frenkel}
\begin{equation}
\rho({\bf r}_o) K({\bf r}_o \rightarrow
{\bf r}_n)=\rho({\bf r}_n) K({\bf r}_n 
\rightarrow
{\bf r}_o),
\end{equation}
where ${\bf r}_o$ and ${\bf r}_n$ represent two
configurations of the system and $K({\bf r}_o \rightarrow
{\bf r}_n)$ is the conditional probability that if the system is at
configuration ${\bf r}_o$ it makes a transition to
${\bf r}_n$.  In many Monte Carlo approaches, the conditional
probability is not known and is replaced by the expression
\begin{equation}
K({\bf r}_o \rightarrow {\bf r}_n) =
T({\bf r}_o \rightarrow {\bf r}_n)
\ {\mathrm acc}({\bf r}_o \rightarrow {\bf r}_n),
\end{equation}
where $T({\bf r}_o \rightarrow {\bf r}_n)$ is called
the trial probability and
${\mathrm acc}({\bf r}_o \rightarrow {\bf r}_n)$ is an acceptance
probability constructed to ensure $K({\bf r}_o \rightarrow
{\bf r}_n)$ satisfies the detailed balance condition.  The trial
probability can be any normalized density function chosen for
convenience.
A common choice for the acceptance probability is given
by\cite{kalos,frenkel}
\begin{equation}
{\mathrm acc}({\bf r}_o \rightarrow {\bf r}_n) =
\min \left [1,\frac{\rho({\bf r}_n)T({\bf r}_n 
\rightarrow {\bf r}_o)}
{\rho({\bf r}_o)T({\bf r}_o 
\rightarrow {\bf r}_n)} \right ].\label{eq:acc}
\end{equation}
The Metropolis method,\cite{metropolis}
obtained from Eq.(\ref{eq:acc}) by choosing 
$T({\bf r}_o \rightarrow {\bf r}_n)$ to be a
uniform distribution of points of width $\Delta$ centered about
${\bf r}_o$, is arguably the most widely used Monte Carlo
method and the basis for all the approaches discussed in the current
work.  The Metropolis method rigorously guarantees a random walk visits
configuration space proportional to a given density function
asymptotically in the limit of an infinite number of steps.  In practice
when configuration space is divided into important
regions separated by significant energy barriers, a low temperature
finite
Metropolis walk can
have prohibitively long equilibration times.

Such problems in attaining
ergodicity in the walk do not occur at temperatures sufficiently high
that the system has significant probability of finding itself in the
barrier regions.  In both the J-walking and parallel tempering
methods, information obtained from an ergodic Metropolis walk at high
temperatures is passed to a low temperature walker periodically to
enable the low temperature walker to overcome the barriers between
separated regions.  In the J-walking method\cite{jwalk1} the trial 
probability 
at inverse temperature $\beta$ is
taken to be a high temperature Boltzmann distribution
\begin{equation}
T({\bf r}_o \rightarrow {\bf r}_n) = Z^{-1}
e^{-\beta_J U({\bf r}_n)}\label{eq:trialj}
\end{equation}
where $\beta_J$ represents the jumping temperature that is sufficiently
high that a Metropolis walk can be assumed to be ergodic.  Introduction
of
Eq.(\ref{eq:trialj}) into Eq.(\ref{eq:acc}) results in the acceptance
probability
\begin{equation}
{\mathrm acc}({\bf r}_o \rightarrow {\bf r}_n) =\min \left\{1, \exp
[-(\beta-\beta_J) (U({\bf r}_n)-U({\bf r}_o))] \right \}.\label{eq:acj}
\end{equation}
In practice at inverse temperature $\beta$ the trial moves are taken
from the Metropolis distribution about 90\% of the time with jumps
attempted using Eq.(\ref{eq:trialj}) about 10\% of the time.  The
jumping configurations are generated with a Metropolis walk at inverse
temperature $\beta_J$, and jump attempts are accepted using
Eq.(\ref{eq:acj}).  The acceptance expression [Eq.(\ref{eq:acj})] is
correct provided the configurations chosen for jumping are a random
representation of the distribution $e^{-\beta_J U({\bf r})}$.  
The Metropolis walk that is used to generate the
configurations at inverse temperature $\beta_J$ is
correlated,\cite{kalos} and
Eq.(\ref{eq:acj}) is inappropriate unless jumps are attempted
sufficiently infrequently to break the correlations.  In practice
Metropolis walks are still correlated after 10 steps, and it is not
possible to use Eq.(\ref{eq:acj}) correctly if jumps are attempted 10\%
of the time.  In J-walking the difficulty with correlations is overcome
in two ways.  In the first method, often called serial
J-walking,\cite{jwalk1} 
a
large set of configurations is stored to an external distribution with
the configurations generated with a Metropolis walk at inverse
temperature $\beta_J$, and configurations stored only after sufficient
steps to break the correlations in the Metropolis walk.  
Additionally, the configurations are chosen from the external
distribution at random.  This external
distribution is made sufficiently large that the probability of ever
choosing the same configuration more than once 
is small.  In this method detailed
balance is strictly satisfied only in the limit that the external
distribution is of infinite size.
In the second method, often called
parallel J-walking,\cite{jwalk2,waljwalk} 
the walks at each temperature are made in tandem on
a parallel machine.  Many processors, randomly initialized, 
are assigned to the jumping temperature, and each processor
at the jumping temperature is used to
donate a high temperature configuration to the low temperature
walk sufficiently infrequently that the correlations in the Metropolis
walk at inverse temperature $\beta_J$ are broken.  In this parallel
method, configurations are never reused, but the acceptance criterion
[Eq.(\ref{eq:acj})] is strictly valid only in the limit of an infinite
set of processors at inverse temperature $\beta_J$.  In practice both
serial and parallel J-walking work well 
for many applications with finite external
distributions or with a finite set of
processors.\cite{jwalk1,jwalk2,waljwalk,jwalk3,strozak,lopez,ni7,jordan1,loplop}

In parallel tempering\cite{temper1,temper2,temper3,temper4,temper5}
configurations from a high temperature walk are
also used to make a low temperature walk ergodic.  In contrast to
J-walking rather than the high temperature walk feeding configurations
to the low temperature walk, the high and low temperature walkers
exchange configurations.  By exchanging configurations detailed balance
is satisfied,
once the Metropolis walks at the two temperatures are
sufficiently long to be in the asymptotic region.  To verify detailed
balance is satisfied by the parallel tempering procedure we let
\begin{equation}
\rho_2({\bf r},{\bf r}') = Z^{-1} e^{-\beta
U({\bf r})}e^{-\beta_J U({\bf r}')}
\end{equation}
be the joint density that the low temperature walker is at configuration
${\bf r}$ and the high temperature walker is at
configuration ${\bf r}'$.  When configurations between the
two walkers are exchanged,
the detailed balance condition is
\begin{equation}
\rho_2({\bf r},{\bf r}') 
K({\bf r} \rightarrow {\bf r}', {\bf r}' 
\rightarrow {\bf r})=
\rho_2({\bf r}',{\bf r}) 
K({\bf r}' \rightarrow {\bf r}, {\bf r}
\rightarrow {\bf r}')
\end{equation}
By solving for the ratio of the conditional transition probabilities
\begin{equation}
\frac{K({\bf r} \rightarrow {\bf r}', {\bf r}' 
\rightarrow {\bf r})}{
K({\bf r}' \rightarrow {\bf r}, {\bf r} 
\rightarrow {\bf r}')}=
\exp
[-(\beta-\beta_J) (U({\bf r}')-U({\bf r}))],
\end{equation}
it is evident that if exchanges are accepted with the same probability
as the acceptance criterion used in J-walking [see Eq.(\ref{eq:acj})], 
detailed balance is satisfied.

Although the basic notions used by both J-walking and parallel tempering
are similar, the organization of a parallel tempering calculation can be
significantly simpler than the organization of a J-walking calculation. 
In parallel tempering
no external distributions are required nor are multiple processors
required at any temperature.  Parallel tempering can be organized in the
same simple way that serial tandem J-walking is organized as discussed
in
the original J-walking reference.\cite{jwalk1} 
 Unlike serial tandem J-walking where
detailed balance can be attained only asymptotically, parallel
tempering satisfies detailed balance directly.
For a
problem as difficult as LJ$_{38}$ where very long simulations are
required, the huge external distributions needed in serial J-walking, or
the large set of jumping processors needed in parallel J-walking,
make the
method prohibitive.  As discussed in Section III, parallel tempering can
be executed for arbitrarily long simulations making the method suitable
at least for LJ$_{38}$.

In the current calculation parallel tempering is used not just to
simulate the system at some low temperature using high temperature
information, but simulations are performed for a series of temperatures.
As is the case for J-walking\cite{jwalk1} 
and as discussed elsewhere for parallel
tempering,\cite{temper4} 
the gaps between adjacent temperatures cannot be chosen
arbitrarily.  Temperature gaps must be chosen so that exchanges are
accepted with sufficient frequency.  If the temperature gap is too
large, the configurations important at the two exchanging temperatures
can be sufficiently dissimilar that no exchanges are ever accepted. 
Preliminary calculations must be performed to explore the temperature
differences needed for acceptable exchange probabilities.  In practice
we have found at least 10\% of attempted exchanges need to be accepted
for the parallel tempering procedure to be useful.  In general the
temperature gaps must be decreased near phase change regions or when the
temperature becomes low.

By exchanging configurations between temperatures, correlations are
introduced at different temperature points.  For example, the average
heat capacities at two temperatures may rise or fall together as each
value fluctuates statistically.  In some cases the values of the heat
capacities or other properties
at two temperatures can be anti-correlated.  The magnitude of
these correlations between temperatures are measured and discussed in
Section III.  As discussed in Section III the correlations between
differing temperatures imply that the statistical fluctuations must be
sufficiently low to ensure any features observed in a calculation as a
function
of temperature are meaningful.

\section{Results}

Forty distinct temperatures have been used in the parallel tempering
simulations of LJ$_{38}$
ranging from $T=0.0143 \varepsilon/k_B$ to $T=0.337 \varepsilon/k_B$. 
The
simulations have been initiated from random configurations of the 38
atoms within a constraining sphere of radius 2.25 $\sigma$.  We have
chosen $R_c=2.25 \sigma$, because we have had difficulties attaining
ergodicity with larger constraining radii.  With large constraining
radii, the system has a significant boiling region at temperatures not
far from the melting region, and it is difficult to execute an ergodic
walk with any method when there is coexistence between liquid-like and
vapor regions.  Constraining radii smaller than $2.25 \sigma$ can induce
significant changes in thermodynamic properties below the temperature of
the melting peak.
Using the
randomly initialized configurations the initialization time to reach the
asymptotic region in the Monte Carlo walk has been found to be long with
about 95 million Metropolis Monte Carlo points followed by 190 
million parallel
tempering Monte Carlo points included in the walk prior to data
accumulation.  This long initiation period can be made significantly
shorter by
initializing each temperature with the structure of the global minimum. 
We have chosen to initialize the system with random configurations to
verify the parallel tempering method is able to equilibrate this system
with no prior knowledge about the structure of the potential surface.
Following this initiation period, $1.3 \times 10^{10}$
points
have been included with data accumulation.  Parallel tempering exchanges
have been attempted every 10 
Monte Carlo passes over the 38 atoms in the cluster.

In an attempt to minimize the
correlations in the data at differing temperatures, an exchange strategy
has been used that includes exchanges between several temperatures.  To
understand this strategy, we let the set of temperatures be put into an
array.  One-half of the exchanges have been attempted between adjacent
temperatures in the array, one-fourth have been attempted between next
near neighboring temperatures, one-eighth between every third
temperature, one-sixteenth between every fourth temperature and
one-thirty second between every fifth temperature in the array.  We have
truncated this procedure at fifth near neighboring temperatures, because
exchanges between temperatures differing by more than fifth neighbors
are accepted with frequencies of less than ten per cent.  
The data presented in this work have been generated using the
procedure outlined above.  In retrospect, 
we have found exchanges are only required
between adjacent temperatures.
We have also
performed the calculations where exchanges are included only between
adjacent temperatures, and we have seen no significance differences
either in the final results or in the correlations between different
temperatures.  
Using the random initializations of
the clusters, after the initialization period the lowest temperature
walks are dominated by configurations well represented by small
amplitude oscillations about the global minimum
structure.

For all data displayed in this work, the error
bars represent two standard deviations of the mean.
The heat capacity, calculated from the
standard fluctuation expression of the energy
\begin{equation}
C_V=k_B\beta^2[\langle E^2 \rangle-\langle E \rangle^2],
\end{equation}
is displayed in the upper panel of Fig. \ref{fig:1}.  In agreement
with the heat
capacity for LJ$_{38}$ reported by Doye {\em et al.},\cite{dwm} the heat
capacity displayed in Fig. \ref{fig:1} has a melting maximum centered at
about
$T=0.166 \varepsilon/k_B$.  In contrast to the results of 
Doye {\em et al.}\cite{dwm} we
find no maximum associated with the solid-solid transition between the
two basins in the potential surface.  Rather, we see a small
change in slope at about $T=0.1 \varepsilon/k_B$.  To characterize 
this region having a change in slope,
in the lower panel of Fig. \ref{fig:1} we present a
graph of $(\partial C_V/\partial T)_{V}$ calculated from the fluctuation
expression
\begin{equation}
\left ( \frac{\partial C_V}{\partial T} \right )_{V} =
-2\frac{C_V}{T} +  \frac{1}{k_B^2T^4}
[\langle E^3 \rangle +2\langle E \rangle^3-
3\langle E^2 \rangle \langle E \rangle ]
\end{equation}
The small low temperature maximum in $(\partial C_V/\partial T)_{V}$
occurs within the slope change region.

To interpret the configurations associated with the various regions of
the heat capacity, we use an order parameter 
nearly identical to the order parameter introduced by Steinhardt,
Nelson and Ronchetti\cite{snr} to distinguish face centered cubic
from icosahedral structures in
liquids and glasses.  The order parameter has been used by Doye {\em et
al.}\cite{dmw} to monitor phase changes in LJ$_{38}$.  The order
parameter $Q_4$ is defined by the equation
\begin{equation}
Q_4=\left ( \frac{4\pi}{9}\sum_{m=-4}^{4} |\overline{Q}_{4,m}|^2 \right
)^{1/2},
\end{equation}
where
\begin{equation}
\overline{Q}_{4,m}=\frac{1}{N_b} \sum_{r_{ij}<r_b}
Y_{4,m}(\theta_{ij},\phi_{ij}).\label{eq:sh}
\end{equation}
To understand the parameters used in Eq.(\ref{eq:sh}), it is helpful to
explain how $\overline{Q}_{4,m}$ is evaluated.  The center of mass of
the
full 38 atom cluster is located and the atom closest to the center of
mass is then identified.  The atom closest to the center of mass plus
the 12 nearest neighbors of that atom define a ``core'' cluster of the
38 atom cluster.  The center of mass of the core cluster is then
calculated.  The summation in Eq.(\ref{eq:sh}) is performed over all
vectors that point from the center of mass of the core cluster to all
$N_b$
bonds formed from the 13 atoms of the core cluster.  A bond is
assumed to be formed between two atoms of the
core cluster if their internuclear separation 
$r_{ij}$ is less than a cut-off parameter
$r_b$, taken to be $r_b=1.39 \sigma$ in this work.  
In Eq.(\ref{eq:sh}) $\theta_{ij}$ and $\phi_{ij}$ are
respectively the polar and azimuthal angles of the vector that points
from the center of mass of the core cluster to the center of each bond,
and $Y_{4,m}(\theta,\phi)$ is a spherical harmonic.  To verify that the
optimal value of $Q_4$ is obtained, the procedure is repeated by
choosing the second closest atom to the center of mass of the whole
cluster to define the core cluster.  The value of $Q_4$ obtained from
this second core cluster is compared with that obtained from
the first core cluster, and the smallest resulting value
of $Q_4$ is taken to be the value of $Q_4$ for the entire cluster.  

In the work of Steinhardt {\em et al.}\cite{snr} 
fewer bonds are included in the
summation appearing in Eq.(\ref{eq:sh}) than in the current work.
In the definition used by Steinhardt {\em et al.},\cite{snr}
the only bonds that contribute to the sum in Eq.(\ref{eq:sh}) are
those involving the central atom of the core cluster.  In the definition
used in this work, at low temperatures the sum includes all the bonds
included by Steinhardt {\em et al.}\cite{snr} 
in addition to vectors that connect
the center of mass of the core cluster with the centers of bonds that
connect atoms at the surface of the core cluster with each other.
For a perfect 
and undistorted icosahedral or
truncated octahedral cluster, the current definition and the definition
of Steinhardt {\em et al.}\cite{snr} 
are identical numerically owing to the
rotational symmetry of the spherical harmonics.  However, for distorted
clusters the two definitions differ numerically.  For perfect,
undistorted icosahedral clusters $Q_4=0$ whereas for perfect,
undistorted
truncated octahedral clusters, $Q_4 \cong 0.19$, and both definitions of
the order parameter are able to distinguish 
configurations from the truncated octahedral
basin and other basins at finite temperatures.  However, we have
found the definition introduced by Steinhardt {\em et al.}\cite{snr} 
is unable to
distinguish structures in the icosahedral basin from liquid-like
structures.  This same issue has been discussed previously by
Lynden-Bell and
Wales.\cite{lbw}  In contrast, we have found that liquid-like structures
have 
larger values of
$Q_4$ than icosahedral structures when the present
definition of $Q_4$ [i.e the definition that includes 
additional bonds in Eq.(\ref{eq:sh})], is used.  
Consequently, 
as discussed shortly, the current definition of $Q_4$
enables an association of each configuration with either the icosahedral
basin, the truncated octahedral basin, or structures that can be
identified as liquid-like.

The average of $Q_4$ as a function of temperature is plotted in the
upper panel of Fig. \ref{fig:2}.  Again the error bars represent two
standard
deviations of the mean.
At the lowest calculated temperatures
$\langle Q_4 \rangle$ is characteristic of the global truncated
octahedral
minimum.  As the temperature is raised to the point where the slope
change begins in the heat capacity, $\langle Q_4 \rangle$ begins to drop
rapidly signifying the onset of transitions between the structures
associated with the global minimum and icosahedral structures.  We then
have the first hint that the slope change
in $C_V$ is associated with a analogue
of a solid-solid transition from the truncated octahedron to icosahedral
structures.

To clarify the transition further, the data plotted in the lower panel
of Fig. \ref{fig:2} represent the probability of observing particular
values of
$Q_4$ as a function of temperature.  The probabilities have been
calculated by tabulating the frequency of observing particular values of
$Q_4$ for each configuration generated in the simulation.  Different
values of $Q_4$ are then assigned to either icosahedral structures
(labeled IC in the graph), truncated octahedral structures (labeled FCC)
or liquid-like structures (labeled LIQ). 
%How the identifications to each kind of structure are made is clarified
below.
By comparing the lower panel of Fig. \ref{fig:2} with the derivative
of the heat capacity plotted in the lower panel of Fig. \ref{fig:1},
it is evident that icosahedral structures
begin to be occupied and the probability of finding truncated octahedral
structures begins to fall when the derivative in the heat capacity
begins to rise.  Equilibrium between the truncated octahedral structures
and the icosahedral structures continues into the melting region, and
truncated octahedral structures only disappear on the high temperature
side of the melting peak of the heat capacity.  Doye {\em et
al.}\cite{dwm}
and Miller {\em et al.}\cite{mdw} 
have generated data analogous to that depicted
in the lower panel of Fig. \ref{fig:2}
using the superposition method, and the data of Miller {\em et
al.}\cite{mdw} 
are in
qualitative agreement with the present data. A more direct comparison
with
the data of these authors can be made by performing periodic quenching
along
the parallel tempering trajectories. We then use an energy criterion
similar to that of Doye {\em et al.}\cite{dwm}
to distinguish the three categories of geometries and to generate the
respective probabilities $P$.
For a given total cluster energy $E$, a truncated
octahedron is associated with $E<-173.26\varepsilon$, icosahedral-based
structures with $-173.26\varepsilon\leq E< -171.6\varepsilon$, and
liquid-like
structures with $E\geq -171.6\varepsilon$. The quenches have been
performed every $10^4$ MC steps for each temperature, and the results of
these
quenches are plotted in Fig. \ref{fig:quenches}. Using the
energy criterion, the behavior we observe is qualitatively
similar to the data of Doye {\em et al.}.\cite{dwm} However, the largest
probability of observing icosahedral structures is found here to be
substantially lower than Doye {\em et al.}\cite{dwm} 
The data accumulated more recently by Miller {\em et al.}\cite{mdw}
using the superposition method include contributions from more
stationary
points than in the previous work of Doye {\em et al.},\cite{dwm} but no
reweighting has been performed. As a result, the distributions of
isomers look quite
different, especially at high temperatures.\cite{dwm}

The assignment of a particular value of $Q_4$ to a structure 
as displayed in Fig. \ref{fig:2}, is made by
an analysis of the probability distribution $P_Q(T,Q_4)$ 
of the order parameter displayed in Figs. \ref{fig:3} and \ref{fig:4}.
Figure \ref{fig:3} is a representation of the three-dimensional surface
of $P_Q(T,Q_4)$ as a function of temperature and order parameter.  A
projection of this surface onto two dimensions is given in Fig.
\ref{fig:4}.
The
probability density in Fig. \ref{fig:4}
is represented by the shading so that the
brighter the area the greater the probability.  The horizontal white
lines in Fig. \ref{fig:4} define the regions of the heat capacity
curve.  The
lowest temperature horizontal 
line represents the temperature at which the slope of
the heat capacity first changes rapidly, 
the middle temperature horizontal line
represents the lowest temperature of the melting peak and the highest
temperature horizontal line represents the end of the melting region. An
additional representation of the data is given in Fig. \ref{fig:5},
where the
probability of observing particular values of $Q_4$ is given as a
function of $Q_4$ at a fixed temperature of $0.14 \varepsilon/k_B$. In
Fig.
\ref{fig:5} three regions are evident for
$P_Q(T=0.14 \varepsilon/k_B,Q_4)$ with $Q_4$ ranging from 0.13 to 0.19.  
Although the presence of three regions
seems to indicate three distinct structures,
all three regions correspond to the truncated octahedral global minimum.
We have verified this assignment by quenching the structures with $Q_4$
ranging from 0.13 to 0.19 to their nearest local minima, and we have
found all such structures quench to the truncated octahedron.  To
explain the three regions, we have found that
there are small distortions of LJ$_{38}$ about the truncated
octahedral structure where
both the energy and $Q_4$ increase together.  
These regions where both the energy
and $Q_4$ increase above $Q_4\cong 0.13$ have low probability and 
account for the oscillations
observed in Figs. \ref{fig:3}--\ref{fig:5}.  In the
lower panel of Fig. \ref{fig:2}, all structures having $Q_4>0.13$ have
been
identified as truncated octahedra.  Quench studies of the broad
region visible in Fig. \ref{fig:4} at
the lowest values of $Q_4$, or equivalently
in the first low $Q_4$ peak in Fig. \ref{fig:5}
find all examined
structures to belong to the icosahedral basin.  
To determine if a given configuration is associated with the icosahedral
basin, one-dimensional cross sectional plots are made from Fig.
\ref{fig:3} at
each temperature used in the calculation. Figure \ref{fig:5} is a
particular
example of such a cross sectional plot.
The maximum present at low $Q_4$ represents the center for
structures in the icosahedral basin.  The next two maxima at higher
$Q_4$
represents the midpoint of the liquid region.
Consequently, in
generating the lower panel of Fig. \ref{fig:2}, all configurations with
$Q_4$
between $Q_4=0$ and the first minimum in Fig. \ref{fig:5}
have been identified as icosahedral structures.  All other values of
$Q_4$, represented by the broad intermediate band in Fig. \ref{fig:4}
(or
the region about the second two maxima in Fig. \ref{fig:5}), 
have been
identified as liquid-like structures.  To make these identifications,
separate cross sections of Fig. \ref{fig:3} must be made at each
temperature.
Of course, it is impossible to
verify that the identification of all values of $Q_4$ with a particular
structure as discussed above would agree with the result of quenching
the structure to its nearest potential minimum. The differences found by
defining icosahedral, truncated octahedral or liquid-like structures
using either an energy criterion or $Q_4$
is clarified by comparing
 Fig. \ref{fig:quenches} and
the lower panel of Fig. \ref{fig:2}. Both definitions
are arbitrary,
and the information carried by the two classification methods 
complement each other.

Figure \ref{fig:4} also provides additional evidence that the peak
in $(\partial
C_V/\partial T)_V$ is associated with the equilibrium between the
truncated octahedral structures and the icosahedral structures.  There
is significant density for both kinds of structures in the region
between the lowest two parallel lines that define the region with the
slope change.  Additionally, both icosahedral structures and truncated
octahedral structures begin to be in equilibrium with each other at the
beginning of the slope change region.  This equilibrium continues to
temperatures above the melting region.

Another identification of the slope change region with a transition
between truncated octahedral and icosahedral forms can be made by
defining $P_R(T,R)dR$ to be the probability that an atom in the cluster
is
found at location $R$ to $R + dR$ from the center of mass of the
cluster at temperature $T$. A
projection of $P_R(T,R)$ onto the $R$ and $T$ plane is
depicted in Fig. \ref{fig:6}.
The solid vertical lines represent the location of
atoms from the center of mass of the truncated octahedral
structure (the lower set of vertical lines), and the lowest energy
icosahedral structure (the upper set of vertical lines).
As in Fig. \ref{fig:4},
increased probability is represented by the lighter shading.  At the
lowest temperatures $P_R(T,R)$ is dominated by contributions from the
truncated octahedron as is evident by comparing the shaded regions with
the lowest set of vertical lines.  As the temperature is increased,
contributions to $P_R(T,R)$ begin to appear from the icosahedral
structures.
 The shaded region at $R=0.45$ does not match any of the vertical lines
shown, but corresponds to atoms in the third lowest energy isomer, which
like the second lowest energy isomer, comes from the icosahedral basin. 
The equilibrium between the icosahedral and truncated octahedral
structures observed in Fig. \ref{fig:6} matches the regions of
temperature
observed in Fig. \ref{fig:4}.

\begin{figure}[tbp]
\caption{$\gamma (T_1,T_2)$ for the heat capacity as defined in
Eq.(\protect\ref{eq:gam}) [with $g=C_V$]
as a function of reduced temperature along two axes.
White shading represents $\gamma=1$ and black represents $\gamma=-1$.
The white diagonal line connecting the lower left hand corner with the
upper right hand corner indicates $T_1=T_2$ so that $\gamma=1$.  The
black areas show anti-correlation from parallel tempering between the
heat capacity calculated at the maximum of the heat capacity and the
center of the change in slope region.}
\label{fig:7}
\end{figure}
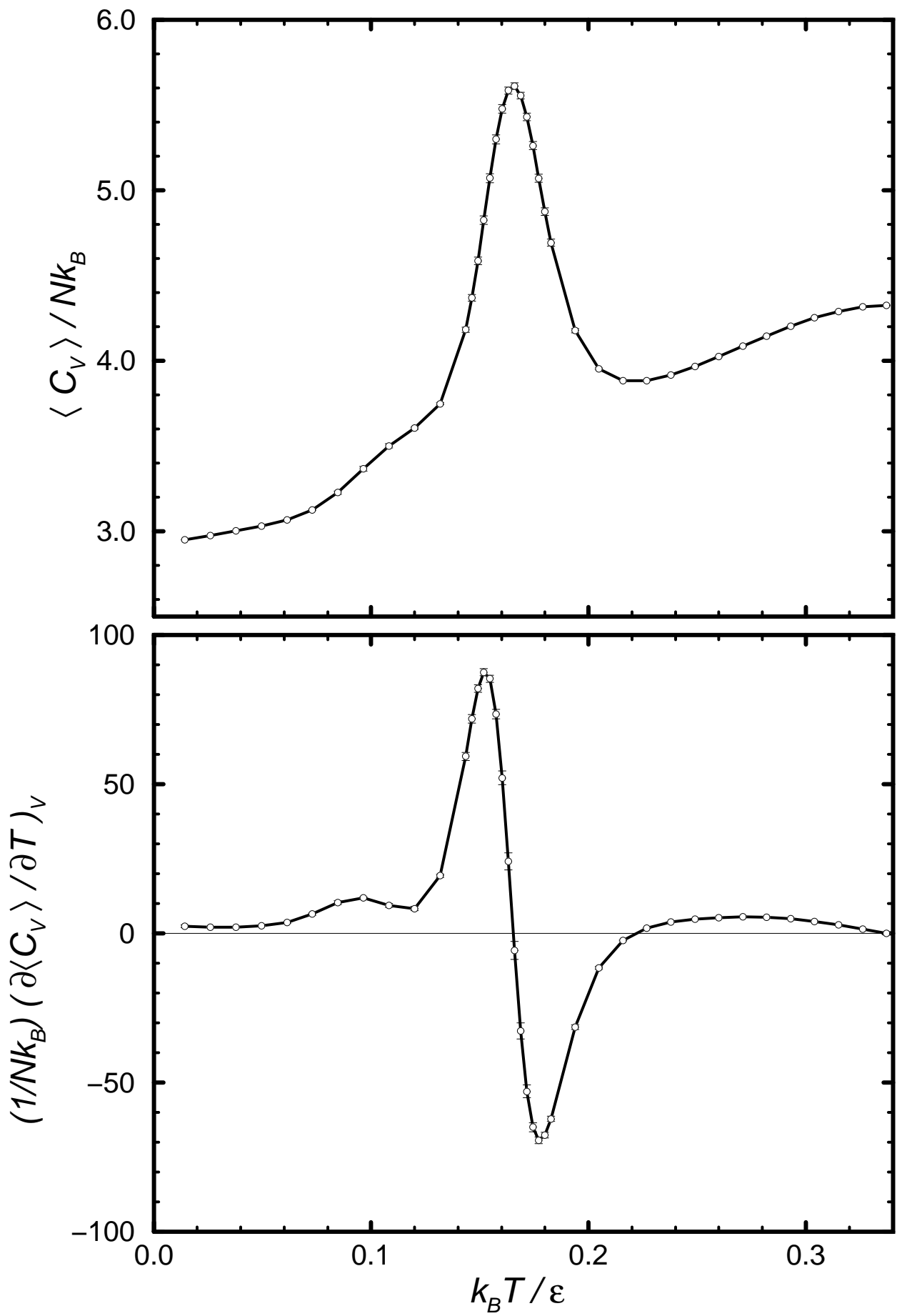
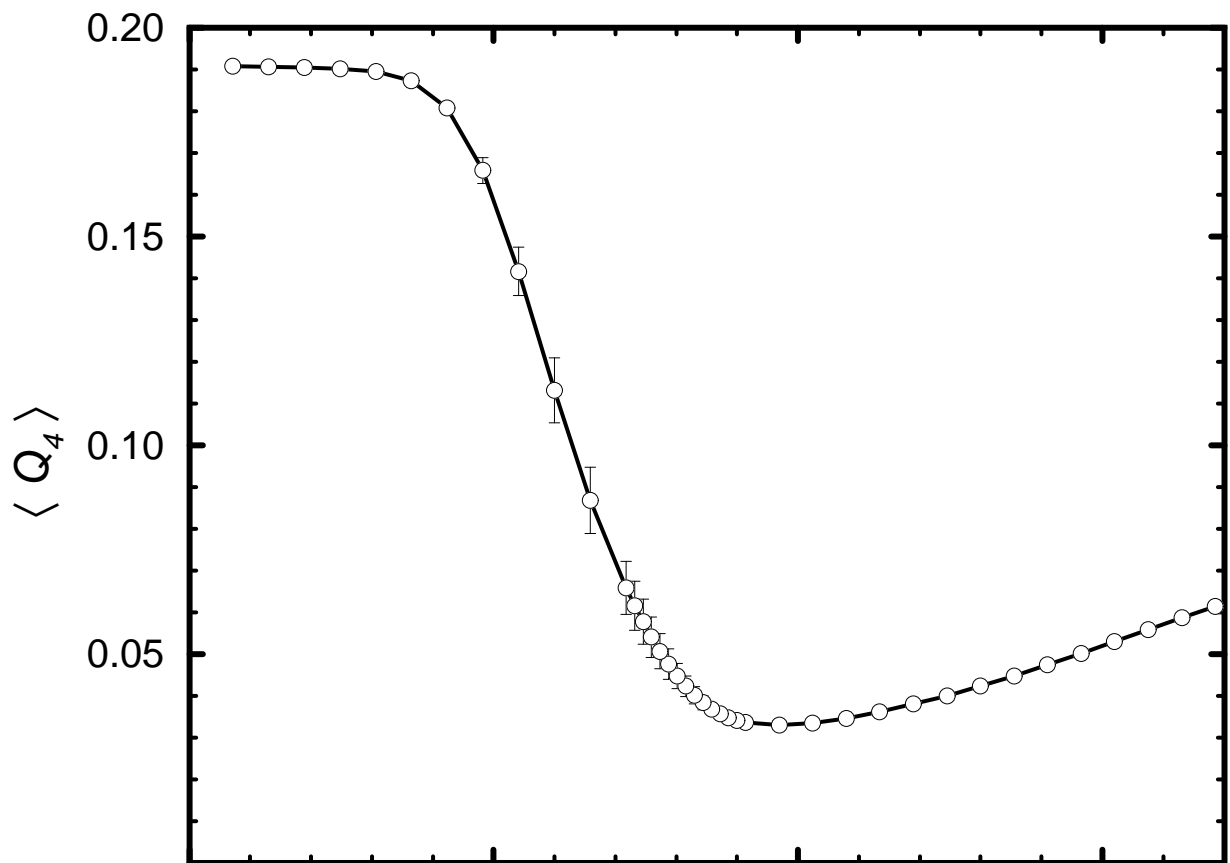
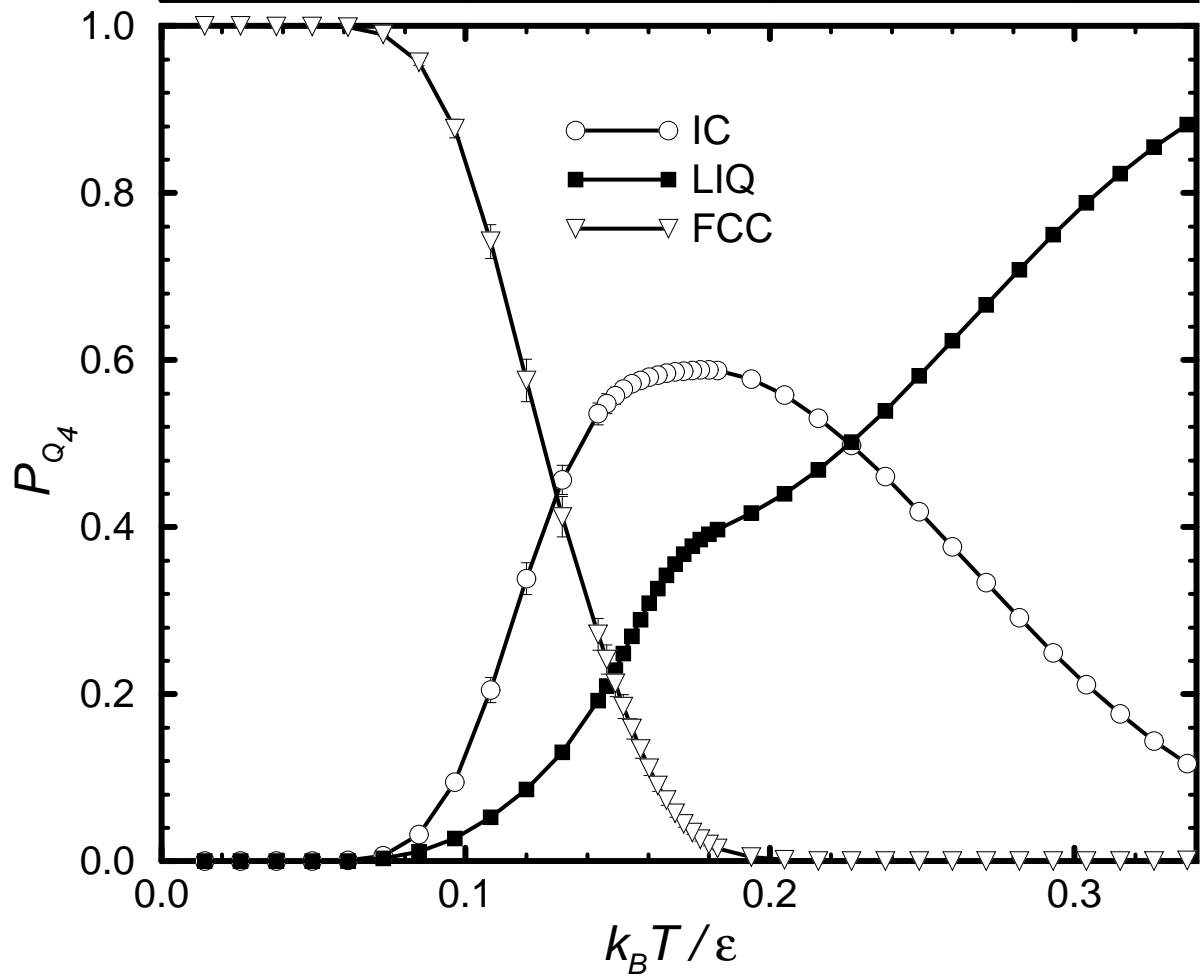
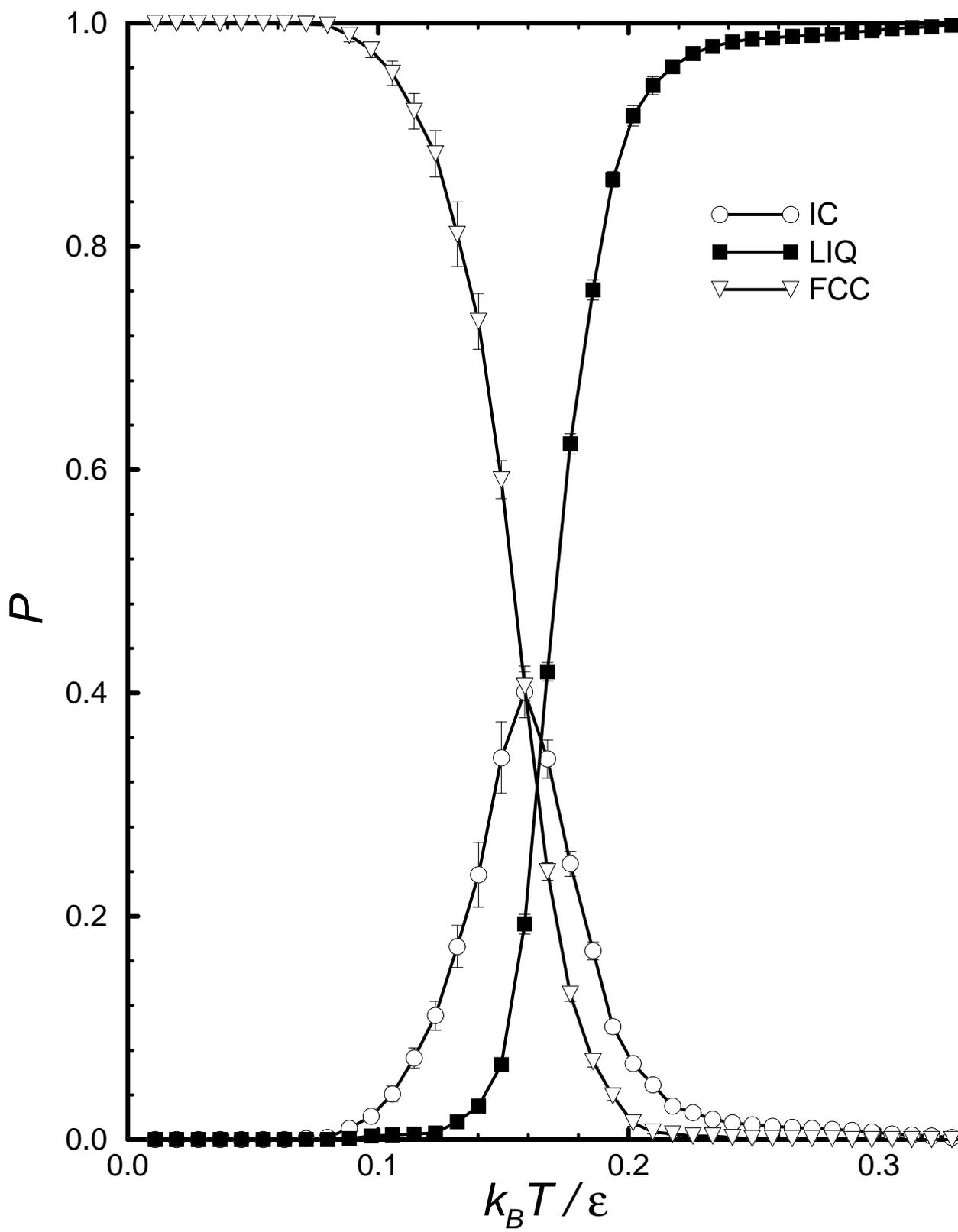
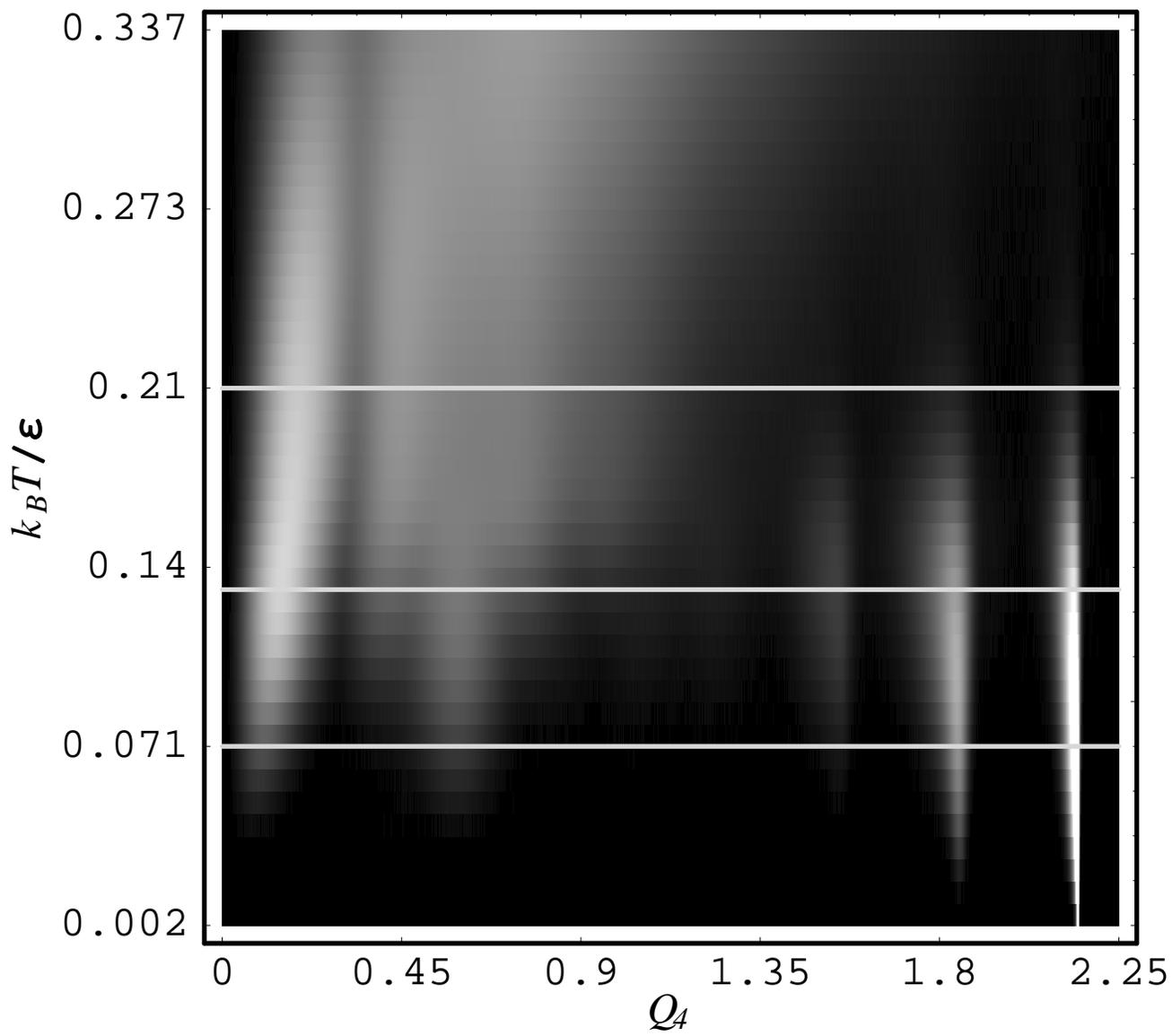
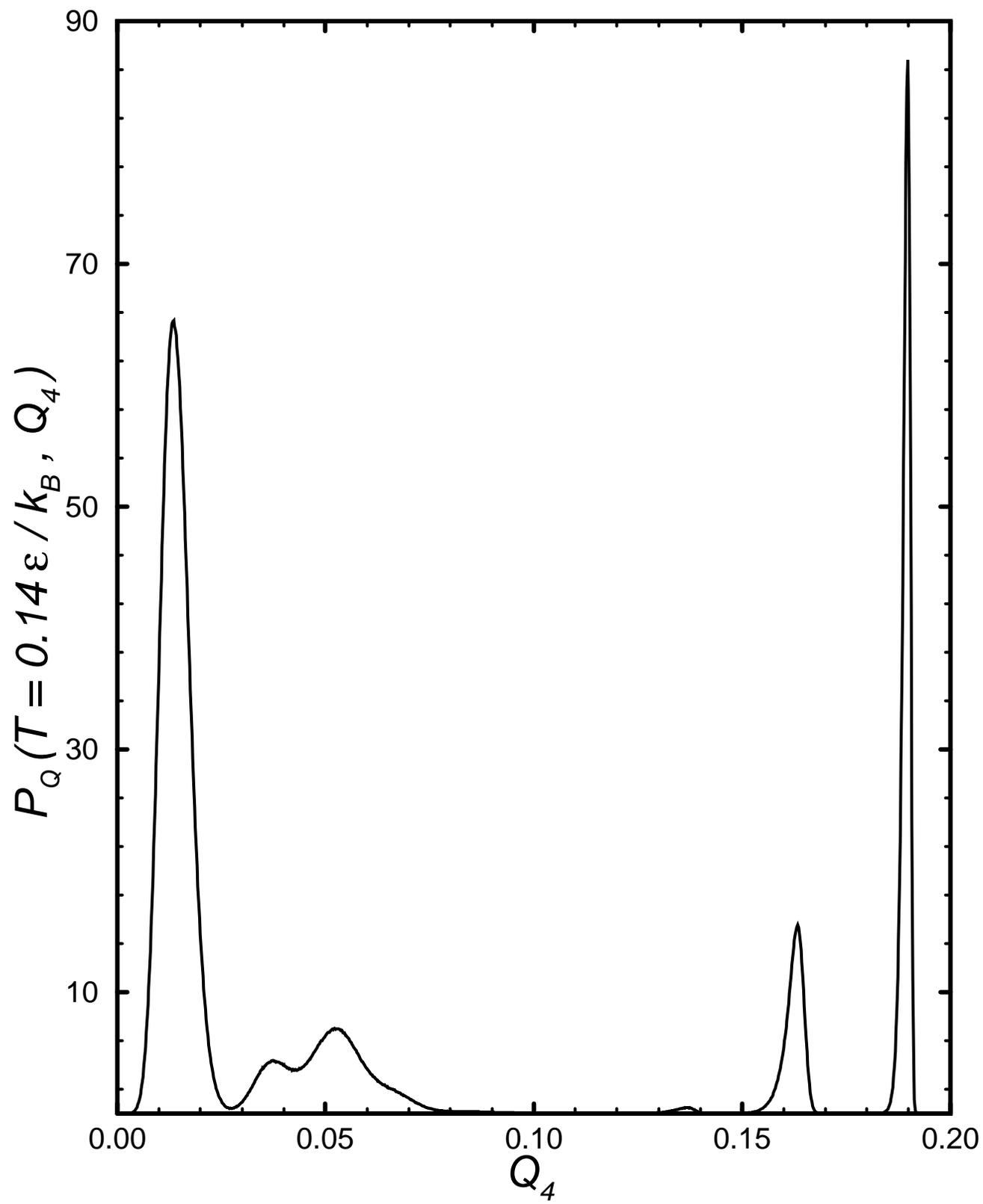
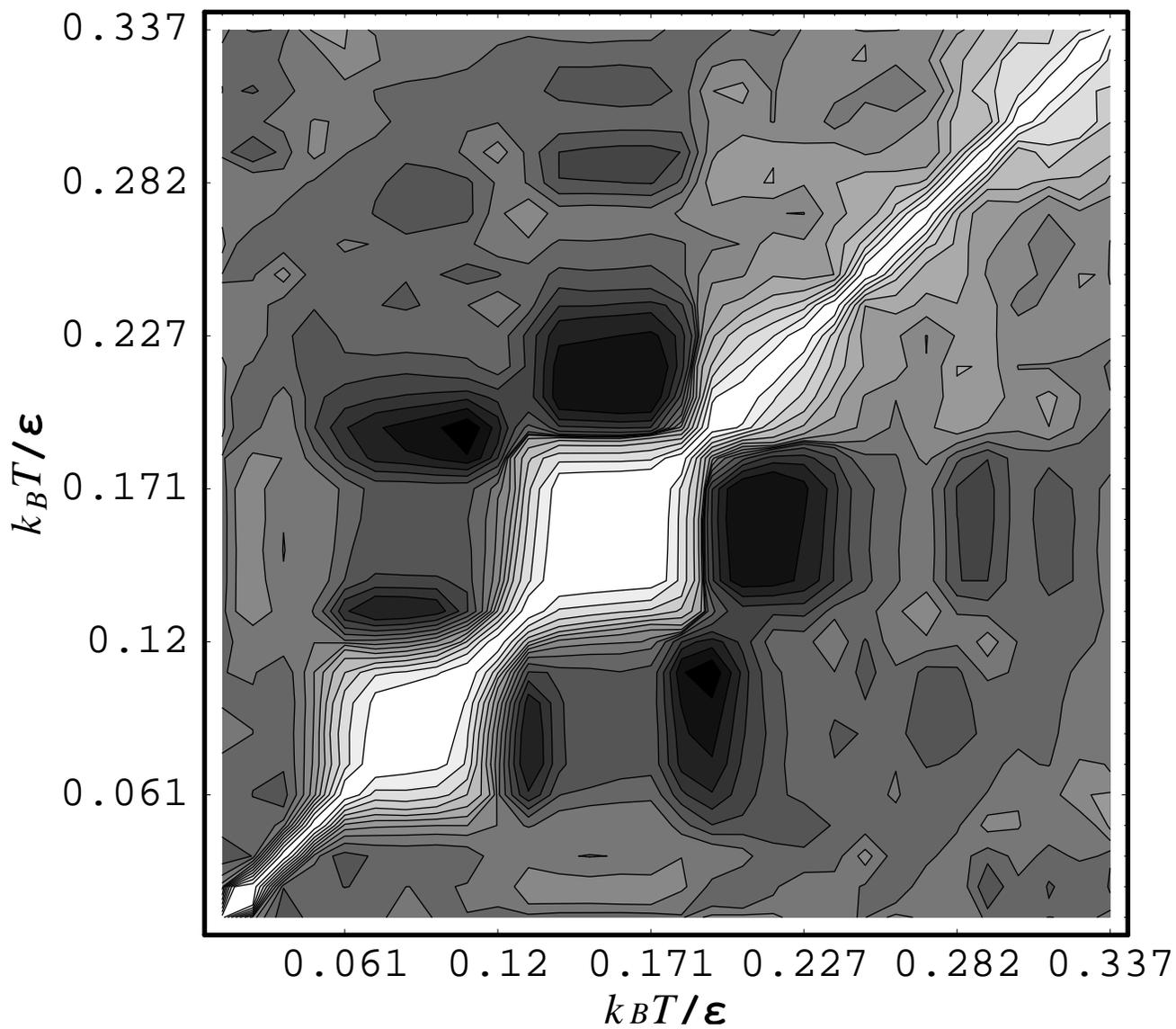
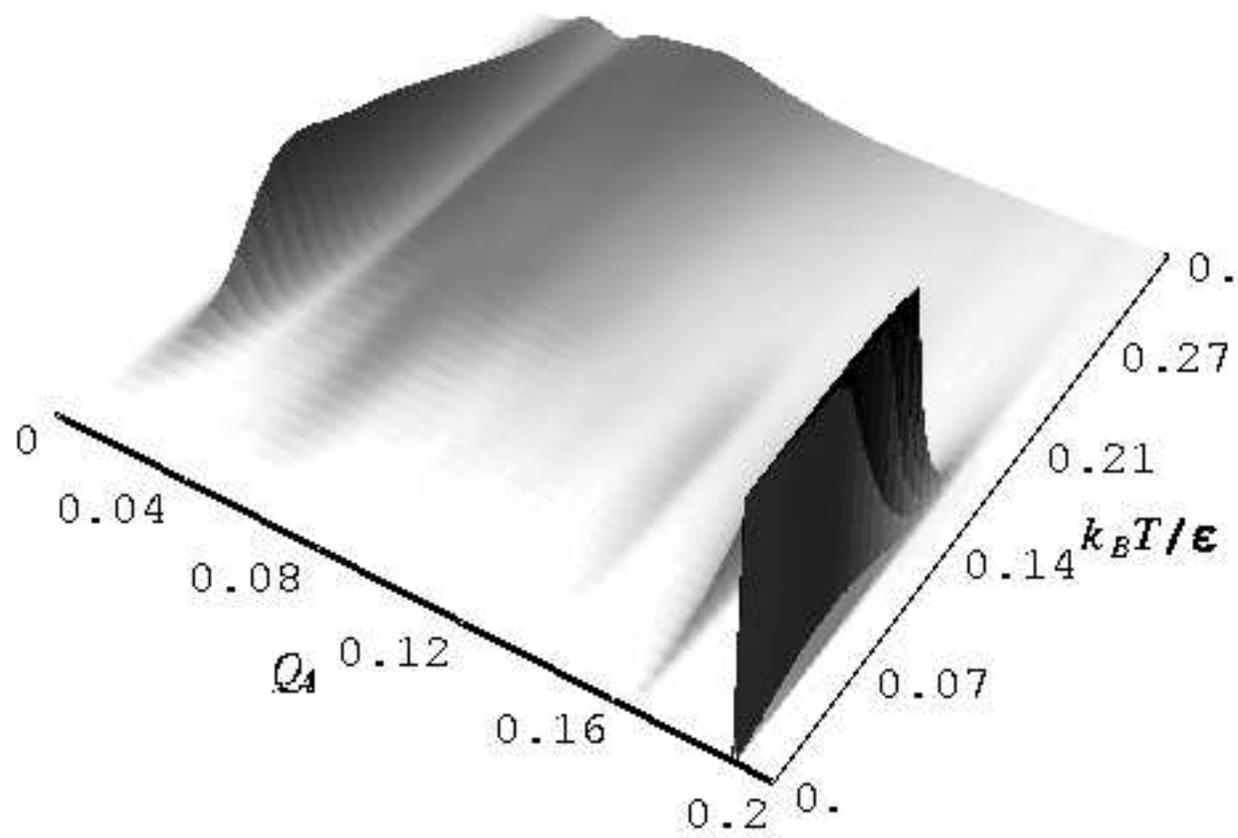
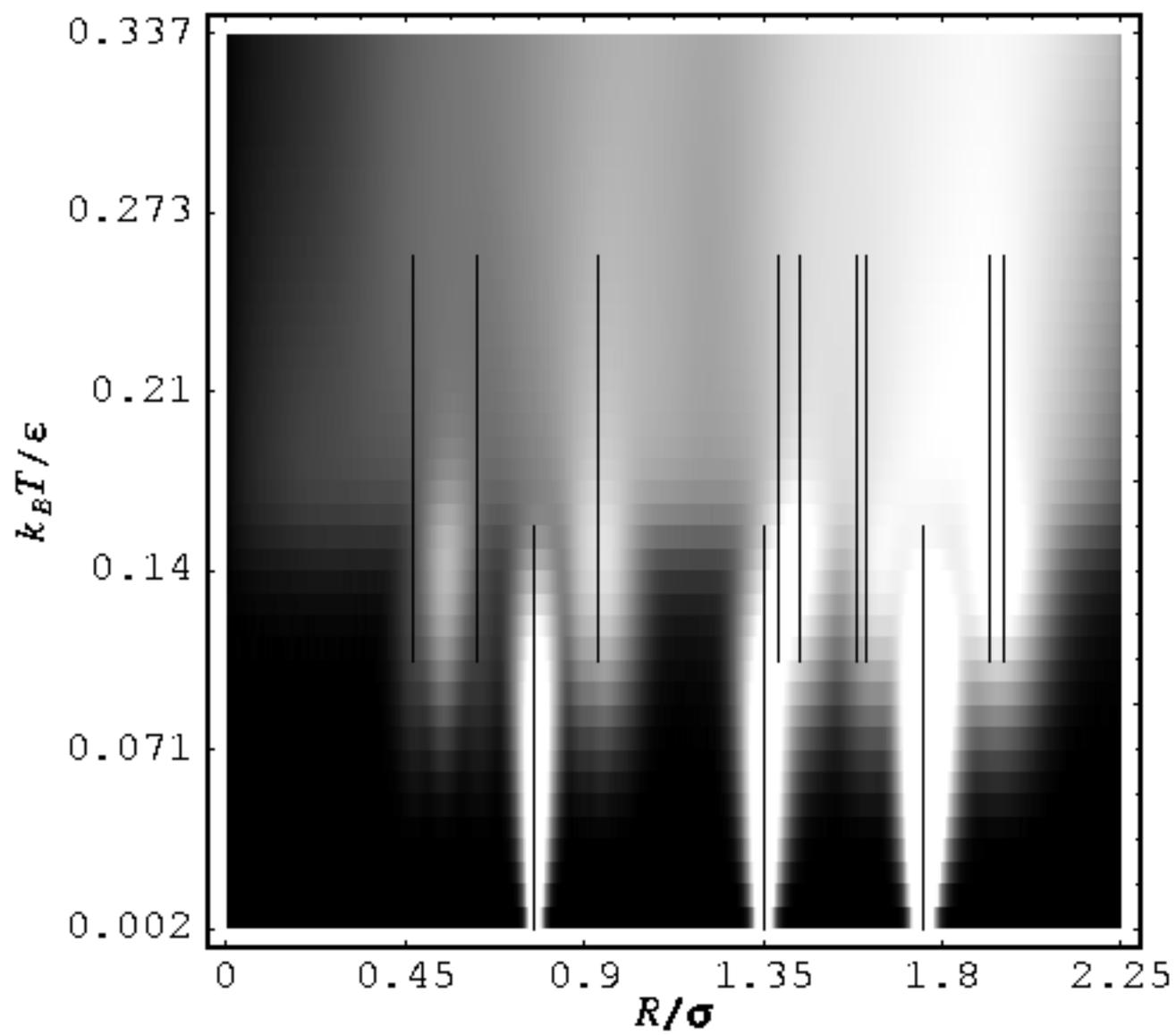

We have mentioned previously that parallel tempering introduces
correlations in the data accumulated at different temperatures, and it
is important to ensure the statistical errors are sufficiently small
that
observed features are real and not artifacts of the correlations.  To
measure these correlations we define a cross temperature correlation
function for some temperature dependent property $g$ by
\begin{equation}
\gamma(T_1,T_2)=\frac{\langle (g(T_1)-\langle g(T_1) \rangle ) (
g(T_2)-\langle g(T_2) \rangle )\rangle}{[\langle (g(T_1)-\langle g(T_1)
\rangle)^2\rangle \langle (g(T_2)-\langle g(T_2)
\rangle)^2\rangle]^{1/2}}.\label{eq:gam}
\end{equation}
A projection of $\gamma(T_1,T_2)$ when $g=C_V$ is given in Fig.
\ref{fig:7}.  In Fig. \ref{fig:7}
white represents $\gamma=1$ and black represents
$\gamma=-1$ with other shadings representing values of $\gamma$ between
these two extremes.  The white diagonal line from the lower left hand
corner to the upper right hand corner represents the case that $T_1=T_2$
so that $\gamma=1$.  The light shaded areas near this diagonal
represent cases where $T_1$ and $T_2$ are adjacent temperatures in the
parallel tempering simulations, and we find $\gamma$ to be only slightly
less than unity.  More striking are the black regions off the diagonal
where $\gamma$ is nearly $-1$.  These black regions correspond to
anti-correlations between results at temperatures near the heat capacity 
maximum in the melting peak and 
temperatures near the center of the slope change region
associated with the transition between icosahedral and truncated
octahedral structures.  These correlations imply the importance of
performing sufficiently long simulations to ensure that statistical
fluctuations of the data are small compared to important features in the
data as a function of temperature.

\section{Conclusions}

Using parallel tempering methods we have successfully performed ergodic
simulations of the equilibrium thermodynamic properties of LJ$_{38}$
in the canonical ensemble.
As discussed by Doye {\em et al.}\cite{dmw} 
the potential surface of this system
is complex with two significant basins; a narrow basin about the global
minimum truncated octahedral structure, and a wide icosahedral basin. 
These two basins are separated both by structure and a large energy
barrier making simulations difficult.  
In agreement with the results of Doye {\em et al.}\cite{dwm} 
we find clear evidence of 
equilibria between structures at the basin of the global
minimum and the icosahedral basin at temperatures below the melting
region.  Unlike previous work we find no heat capacity maximum
associated with this transition, but rather a region with a change
in the slope of the heat capacity as a function of temperature.

We have found parallel tempering to be successful with this system, and
have noted correlations in our data at different temperatures when the
parallel tempering method is used.  These correlations imply the need to
perform long simulations so that the statistical errors are sufficiently
small that the correlations do not introduce artificial conclusions.

We believe that the methods used in this work could be applied to a 
variety of other systems
including clusters of complexity comparable to LJ$_{38}$.
For instance, the 75-atom Lennard-Jones cluster is known to share many
features with the 38-atom cluster investigated here. LJ$_{75}$ is 
also characterized by a double funnel
energy landscape, one funnel being associated with icosahedral
structures, 
and the other
funnel being associated with the decahedral global minimum. 
The landscape of LJ$_{75}$ has been recently
investigated by Doye, Miller  and Wales\cite{dmw1} who have used
$Q_6$ as the order
parameter.
In another paper,\cite{wd} Wales and Doye have predicted that
the temperature where the decahedral/icosahedral equilibrium takes 
place should be
close to $0.09\varepsilon/k_B$. This prediction is made by using the 
superposition
method, but no caloric curves have yet been reported for LJ$_{75}$. 
The parallel
tempering Monte Carlo method can be expected to work well
for LJ$_{75}$, and
such a parallel tempering study would be another 
good test case for theoretical methods discussed in this work.

A useful enhancement of parallel tempering Monte Carlo is the use of
multiple histogram methods\cite{labwhet,swen}
that enables the calculation of thermodynamic
functions in both the canonical and microcanonical ensembles 
by the calculation of the microcanonical entropy.  In practice the
multiple histogram
method requires the generation of histograms of the potential energy at
a set of temperatures such that there is appreciable overlap of the
potential energy distributions at adjacent temperatures.  This overlap
requirement is identical to the choice of temperatures needed in
parallel tempering.

In performing simulations on
LJ$_{38}$ we have tried other methods to reduce ergodicity errors,
and we close this section by summarizing the difficulties we have
encountered with these alternate methods.  It is important to recognize
that the parallel tempering simulations include in excess of $10^{10}$
Monte Carlo points, and most of our experience with these alternate
methods have come from significantly shorter simulations.  Our ability
to include this large number of Monte Carlo points with parallel
tempering is an important reason why we feel parallel tempering is so
useful.

>From experience with other smaller and simpler clusters,
for a J-walking simulation to include $10^{10}$ points, an external
distribution containing at least $10^9$ points is required to prevent
oversampling of the distribution.  Such a large distribution is
prohibitive with current computer technology.  Our J-walking simulations
containing about $10^7$ Monte Carlo points have resulted in data that
have not been internally reproducible, and data that are not in good
agreement 
with the parallel tempering data.  Many long J-walking simulations with
configurations initiated at random only have icosahedral structures
at the lowest calculated temperatures.
To stabilize the J-walking method with respect to the inclusion of
truncated octahedral structures at low temperatures, we
have attempted to generate distributions using the modified potential
energy function $U_m({\bf r},\lambda)=U({\bf r})
-\lambda Q_4$.  In this modified potential $\lambda$ is a parameter
chosen to deepen the octahedral basin without significantly distorting
the cluster.
While this modified potential has led to more stable
results than J-walking using the bare potential, the results with $10^8$
Monte Carlo points have not been reproducible in detail.  The
application of Tsallis distributions\cite{straub} 
has not improved this situation.

We have also tried to apply the multicanonical J-walking approach
recently introduced by Xu and Berne.\cite{berne} 
While this multicanonical approach
has been shown to improve the original J-walking strategy for other
cluster systems, in the case of LJ$_{38}$ the iterations needed to
produce the external multicanonical distribution have not produced
truncated octahedral structures.  The iterations have produced external
distributions having either liquid-like structures or structures from
the icosahedral basin.  The multicanonical distribution is known to
have deficiencies at low energies, and this low energy
difficulty appears to be problematic for LJ$_{38}$. We have attempted
to solve these deficiencies by including prior information about the
thermodynamics of the system. In this attempt we have chosen the 
multicanonical weight to be
$w_{\mathrm mu}(U)=\exp [-S_{\mathrm PT}(U)]$ where
$S_{\mathrm PT}(U)$ is the microcanonical entropy extracted from a
multihistogram analysis\cite{labwhet,swen} of a
parallel tempering Monte Carlo simulation.  In several
attempts using this approach we have not observed either the
truncated octahedral structure nor structures from the icosahedral basin
with significant probability.  The multicanonical distribution so
generated is dominated by liquid-like structures, and the distribution
appears to be incapable of capturing the solid-to-solid transition that
leads to the low temperature peak in $(\partial C_V/\partial T)_{V}$. 
Whether there are other approaches to generate a
multicanonical distribution that are more successful in capturing low
temperature behaviors is unknown to us.

Much
insight about phase change behaviors can be obtained from simulations in
the microcanonical ensemble or using molecular dynamics methods.  For
example, the van der Waals loops observed in LJ$_{55}$\cite{labwhet} 
complement the
interpretation of the canonical caloric curves.  In the next
paper\cite{paper2} 
we
present parallel tempering results for LJ$_{38}$ using both
molecular dynamics and microcanonical Monte Carlo methods.

\section*{Acknowledgments}

Some of this work has been 
motivated by the attendance of two of us (DLF and FC)
at a recent CECAM meeting on `Overcoming broken ergodicity in
simulations of
                    condensed matter systems.' 
We would like to thank CECAM, J.E. Straub
and B. Smit who organized the meeting, and those who attended the
workshop for stimulating discussions, particularly on the connections
between J-walking and parallel tempering.
Two of us (DLF and JPN) would also like to thank Professor M.P.
Nightingale
for helpful discussions concerning the parallel tempering method.
This work has been supported in part by the National Science Foundation
under
grant numbers CHE-9714970 and CDA-9724347.
This research has 
been supported in
part by the Phillips Laboratory, Air Force Material Command, USAF,
through the use of the MHPCC under cooperative agreement number
F29601-93-0001.  The views and conclusions contained in this document
are those of the authors and should not be interpreted as necessarily
representing the official policies or endorsements, either expressed or
implied, of Phillips Laboratory or the U.S. Government.

\newpage

\begin{figure}[tbp]
\caption{The heat capacity $C_V$ per particle of LJ$_{38}$
 in units of $k_B$ (upper panel)
and $(\partial C_V/\partial T)_V$ per particle 
(lower panel) as a function of reduced
temperature.  The small low temperature maximum in the derivative
associated with a change in slope of the heat capacity identifies the
transition region between the truncated octahedral basin and the
icosahedral basin.  The large heat capacity peak identifies the melting
region.}
\label{fig:1}
\end{figure}

\begin{figure}[tbp]
\caption{The expectation value of the order parameter (upper
panel) and the order parameter probability distribution (lower panel) as
a function of reduced temperature.  In the lower panel FCC labels the
truncated octahedron, IC labels structures from the icosahedral basin
and LIQ labels structures from the liquid region.  The transition
between FCC and IC occurs at the same temperature as the 
low temperature peak in
$(\partial C_V/\partial T)_V$ in Fig. \protect\ref{fig:1}.}
\label{fig:2}
\end{figure}

\begin{figure}[tbp]
\caption{The probability distributions of observing different structures
as a function of temperature using the energy criterion.
The labels are the same as those defined in the lower panel of
Fig. \protect\ref{fig:2}, and the data complements the interpretation
of the lower panel of Fig. \ref{fig:2}}
\label{fig:quenches}
\end{figure}

\begin{figure}[tbp]
\caption{The probability of observing configurations with
particular values of $Q_4$ with $Q_4$ displayed along one axis and the
reduced temperature displayed along the other axis.  The large peak at
low temperatures comes from the truncated octahedral structures and the
broad region with small $Q_4$ at intermediate temperatures represents
structures in the icosahedral basin.}
\label{fig:3}
\end{figure}

\begin{figure}[tbp]
\caption{A projection of Fig. \protect\ref{fig:3} onto the $T$-$Q_4$
plane.  
The probability is measured by the shading with increasing
probability represented by lighter shading. The
lowest temperature horizontal white line represents the temperature at
which transitions between the icosahedral and lowest energy basins
begin.  
The second lowest temperature horizontal white line represents
the beginning of the melting region, and the highest temperature
horizontal white line represents the end of the melting peak of the heat
capacity.  The coexistence of icosahedral and octahedral structures
continues into the melting region.}
\label{fig:4}
\end{figure}

\begin{figure}[tbp]
\caption{The probability of observing configurations with
particular values of $Q_4$ as a function of $Q_4$ at $T=0.14
\varepsilon/k_B$. 
The region from $Q_4=0$ through the first maximum to the first minimum
defines the icosahedral basin at $T=0.14 \varepsilon/k_B$, the region
from the
first minimum to the third defines liquid-like structures, and the
region about the three maxima having the highest values of $Q_4$ define
the truncated octahedral basin.  The oscillations in the truncated
octahedral basin arise from distorted structures of low probability
where both the energy and $Q_4$ rise together.}
\label{fig:5}
\end{figure}

\begin{figure}[tbp]
\caption{The projected probability of observing particles a
distance $R$ from the center of mass of LJ$_{38}$ as a function of $R$
and reduced temperature. As in Fig. \protect\ref{fig:4}
increased probability is represented by the lightest shading.
The lower vertical lines represent the
location of atoms in the fully relaxed truncated octahedron and the
upper vertical lines represent the location of atoms in the fully
relaxed icosahedral structure that is lowest in energy.  Equilibrium
between the icosahedral and octahedral forms are observed in the same
temperature range as found in Figs. \protect\ref{fig:2} and
\protect\ref{fig:4}.}
\label{fig:6}
\end{figure}

\end{document}